\documentstyle[stwol,epsfig]{article}

\newcommand{\AmS}{{\protect\the\textfont2
  A\kern-.1667em\lower.5ex\hbox{M}\kern-.125emS}}

\input epsf
\def\D0{D\O}
\def\etmisv {\mbox{${\hbox{${\vec E}$\kern -0.65em\lower-.1ex\hbox{/}}}_T$}}
\def\etmis  {\mbox{${\hbox{$E$\kern -0.65em\lower-.1ex\hbox{/}}}_T$}}
\def\pbarp  {\mbox{$\overline{p}p$} }
\def\pdf    {parton distribution function }

\def\ifmath#1{\relax\ifmmode #1\else $#1$\fi}%
\def\GeV{\ifmmode {\mathrm{ Ge\kern -0.1em V}}\else
                   \textrm{Ge\kern -0.1em V}\fi}%
\def\MeV{\ifmmode {\mathrm{ Me\kern -0.1em V}}\else
                   \textrm{Me\kern -0.1em V}\fi}%
\def\keV{\ifmmode {\mathrm{ ke\kern -0.1em V}}\else
                   \textrm{ke\kern -0.1em V}\fi}%
\def\eV{\ifmmode  {\mathrm{ e\kern -0.1em V}}\else
                   \textrm{e\kern -0.1em V}\fi}%
\def\GeVcc{\ifmmode {\mathrm{ \GeV/c^2}}\else
                   \textrm{Ge\kern -0.1em V/c$^2$}\fi}%

\def\pbarp              {\mbox{$\overline{p}p$ }}
\def\ppbar              {\mbox{$\overline{p}p$ }}
\def\ttbar              {\mbox{$t\overline{t}$ }}

\newcommand{\AFB}       {A_{\mathrm{FB}}}
\newcommand{\ee}        {\mbox{$e^+e^-$}}
\newcommand{\qq}        {\mbox{$q\overline q$}}
\newcommand{\uu}        {\mbox{$u\overline u$}}
\newcommand{\dd}        {\mbox{$d\overline d$}}

\newcommand{\SM}        {\mbox{SM}}

\begin{document}

\begin{titlepage}
\rightline{\vbox{\halign{&#\hfil\cr
&Fermilab-Conf-97/296\cr
&August 1997\cr}}}
\vspace{1in}

\renewcommand{\thefootnote}{\fnsymbol{footnote}}
\setcounter{footnote}{2}

\begin{center}

{\Large\bf
    Electroweak Measurements from Hadron Colliders\footnote{\normalsize
    Invited talk given at the
    {\it  XVI International Workshop on Weak Interactions and Neutrinos }, 
    Capri, Italy, June 22 -- 28, 1997.}
}

\vskip 2.0cm
\normalsize
{\large Marcel Demarteau}
\vskip .3cm
Fermilab \\
Batavia, IL 60510, USA\\
\vskip 3cm

\end{center}

\begin{abstract}

A review of recent electroweak results from hadron colliders
is given. Properties of the $W^\pm$ and $Z^0$ gauge bosons
using final states containing electrons and muons based on large 
integrated luminosities are presented. The emphasis is placed on the 
measurement of the mass of the $W$ boson and the measurement of
trilinear gauge boson couplings. 

\end{abstract}

\setcounter{footnote}{0}
\renewcommand{\thefootnote}{\arabic{footnote}} 
\end{titlepage}

\title{Electroweak Results from Hadron Colliders}

\author{Marcel Demarteau 
	\address{ {\it Fermilab, \ 
        P.O. Box 500, \ Batavia, \ IL 60510 }}%
%         \thanks{Work supported by the U.S. Dept. of Energy under 
%                 contract DEAC02-76CHO3000 }
       }

% \begin{abstract} 

\twocolumn[\maketitle\abstracts{
A review of recent electroweak results from hadron colliders
is given. Properties of the $W^\pm$ and $Z^0$ gauge bosons
using final states containing electrons and muons based on large 
integrated luminosities are presented. The emphasis is placed on the 
measurement of the mass of the $W$ boson and the measurement of
trilinear gauge boson couplings. 
}]
% \end{abstract}

\section{Introduction }

The standard model of electroweak interactions
(SM) has taken a very prominent position in today's description of 
experimental results. 
Per\-haps the most compelling reason for this state of affairs is that 
the experimental results have reached a level of precision which require 
a comparison with theory beyond the Born calculations, which the $\SM$ 
is able to provide. 
It is widely anticipated, though, that the $\SM$ is just an approximate 
theory and should eventually be replaced by a more complete and
fundamental description of the underlying forces in nature. 
Since the highest center of mass energies are reached at
hadron colliders, notably the Tevatron, the measurements at this 
accelerator provide natural tools 
to probe the $\SM$ at the highest energy scale. 

In this summary the most recent electroweak results 
from the multi-purpose detectors CDF and \D0 
operating at the Fermilab Tevatron \pbarp Collider will be 
described. 
The \D0 detector has a non-magnetic inner 
tracking system, compact, hermetic, uranium liquid-argon calorimetry and 
an extensive muon system.
The CDF detector has a magnetic central detector, scintillator based 
calorimetry and a central muon system. 
During the 1992-1993 run, generally called Run~Ia, 
the CDF and \D0 experiments have collected 
$\sim$20~pb$^{-1}$ and $\sim$15~pb$^{-1}$ of data, respectively. For the 
1994-1995 run (Run~Ib) both experiments have collected 
$\sim$90~pb$^{-1}$ of data. 
First, results on inclusive and differential $W$ and $Z$ production 
cross sections % and some derived results 
are presented. 
The $W$ mass measurement is then described with its do\-mi\-nant 
uncertainties. 
In the last section triple gauge boson interactions 
are discussed.

\section{IVB Production Cross Sections }

In \pbarp collisions intermediate vector bosons are produced predominantly 
by quark-antiquark annihilation. At $\sqrt{s} = 1.8$~TeV 
sea-sea interactions contribute approximately 
20\% to the total cross section. The leptonic decay modes of the 
$W$ and $Z$-bosons are easily detected because of their characteristic decay 
signatures: for a $W$ decay a high $p_T$ lepton accompanied by 
large missing transverse energy 
(\etmis), 
indicating the presence of a neutrino, and two high 
$p_T$ leptons for $Z$-decays. 
The measurement of the $W$ and $Z$ production cross sections probes the 
$\SM$ of electroweak and strong interactions and provides insight 
in the structure of the proton. 
% With the large increase in integrated luminosity the new measurements 
% are anticipated to have a significantly improved precision. 
A persistent uncertainty on any cross section measurement at a \pbarp 
collider, however, is the large uncertainty on the integrated luminosity 
due to the uncertainty on the effective total \pbarp cross section 
seen by the detectors. 
This uncertainty cancels completely in the ratio of the $W$ and $Z$ 
production cross sections, a quantity that 
% Since the production cross section measured is the product 
% of the total production cross section and the leptonic branching ratio, 
% the ratio of the production cross sections can be used to extract the width 
can be used to extract the width 
of the $W$-boson, $\Gamma_W$. 
The measurement of the individual cross sections is thus geared towards 
maximizing the cancellation of the different uncertainties in the ratio 
of the two cross section measurements.

\begin{table}[ht]
\begin{center}
{\footnotesize 
\begin{tabular}{l|c|c||c} \hline\hline
                & \multicolumn{2}{c||}{ \D0  }
                & \multicolumn{1}{c}{   CDF  }  \\ \hline
                & $e$   & $\mu$      
                & $e$             \\ \hline 
$W$ Candidates      & 59579          
                    & 4472 
                    & 13796       \\    
A$_W$ (\%)          & 43.4 $\pm$ 1.5 & 20.1 $\pm$ 0.7 
                    & 34.2 $\pm$ 0.8  \\
$\epsilon_W$ (\%)   & 70.0 $\pm$ 1.2 & 24.7 $\pm$ 1.5 
                    & 72.0 $\pm$ 1.2  \\
Bkg $W$ (\%)        &  8.1 $\pm$ 0.9 & 18.6 $\pm$ 2.1 
                    & 14.1 $\pm$ 1.3  \\ \hline 
$\int {\cal L}$ (pb$^{-1}$)
                    & 75.9 $\pm$ 6.4 & 32.0 $\pm$ 2.7 
                    & 19.7 $\pm$ 0.7  \\ \hline 
$Z$ Candidates      & 5702           
                    & 173 
                    & 1312        \\ 
A$_Z$ (\%)          & 34.2 $\pm$ 0.5 &  5.7 $\pm$ 0.5 
                    & 40.9 $\pm$ 0.5  \\
$\epsilon_Z$ (\%)   & 75.9 $\pm$ 1.2 & 43.2 $\pm$ 3.0 
                    & 69.6 $\pm$ 1.7  \\
Bkg $Z$ (\%)        &  4.8 $\pm$ 0.5 &  8.0 $\pm$ 2.1 
                    &  1.6 $\pm$ 0.7  \\
$\int {\cal L}$ (pb$^{-1}$)
                    & 89.1 $\pm$ 7.5 & 32.0 $\pm$ 2.7 
                    & 19.7 $\pm$ 0.7  \\ \hline \hline
\end{tabular}
}
\end{center}
\caption[]{Analysis results for the $W$ and $Z$-production cross section 
           measurement for CDF and preliminary results for \D0. 
           A$_V$, $\epsilon_V$ and Bkg stand for acceptance, detection 
           efficiency and Bkg, respectively, for vector boson $V$. }
\label{table:xsec}
\end{table}

The event selection for $W$-bosons requires an isolated lepton with 
transverse momentum $p_T > 25~(20)$~GeV and $\etmis > 25~(20)$~GeV 
for \D0 (CDF). Leptonic 
decays of $Z$-bosons are selected by imposing the same lepton quality 
and kinematic cuts 
on one lepton, and looser requirements on the second lepton. 
Table~\ref{table:xsec} lists the kinematic and geometric acceptance
(A$_V$), trigger and event selection efficiency ($\epsilon_V$) and 
background (Bkg) for the electron and 
muon decay channel for the two experiments ($V=W/Z$) 
\cite{d0_cdf_xsec_1a}.

\begin{table}[h]
\begin{center}
{\footnotesize 
\begin{tabular}{||l|l|l||} \hline\hline
      & \multicolumn{1}{c|}{ 
        $\sigma_W \cdot B (W\rightarrow \ell\nu)$  }
      & \multicolumn{1}{c||}{ 
        $\sigma_Z \cdot B (Z\rightarrow \ell\ell)$  }  \\ \hline
                        &                                   &   \\
\D0 (e)                 & 2.38  $\pm$ 0.01  $\pm$ 0.22 
                        & 0.235 $\pm$ 0.003 $\pm$ 0.021         \\
\D0 ($\mu$)             & 2.28  $\pm$ 0.04  $\pm$ 0.25 
                        & 0.202 $\pm$ 0.016 $\pm$ 0.026         \\
CDF (e)                 & 2.49  $\pm$ 0.02 $\pm$ 0.12 
                        & 0.231 $\pm$ 0.006 $\pm$ 0.011         \\
% CDF ($\mu$)             & 2.48  $\pm$ 0.031 $\pm$ 0.16
%                         & 0.203 $\pm$ 0.010 $\pm$ 0.012         \\ 
&&\\    \hline\hline
\end{tabular}
}
\end{center}
\caption[]{Preliminary \D0 results on the measured cross section times 
	   branching ratio in nb for $W$ and $Z$ production 
	   from the 1994-1995 run based on an integrated luminosity of 
	   89.1 (32.0) pb$^{-1}$ for the electron (muon) decay, 
	   and published CDF results from the 1992-1993 data based on 
	   an integrated luminosity of 
	   19.7~pb$^{-1}$.  }
\label{table:xsec-result}
\end{table}

\begin{figure}[t]
    \epsfxsize = 8.0cm
    \centerline{\epsffile{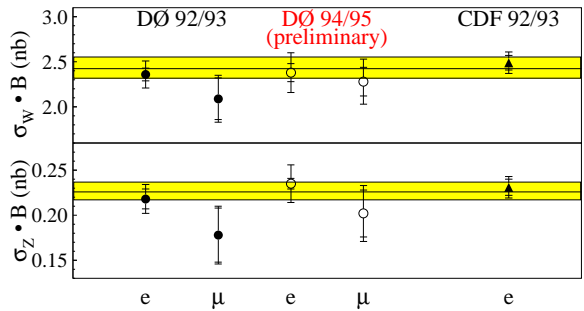}}
\caption{Measurements of the $W$ and $Z$ inclusive cross section compared 
         with the theoretical prediction using the CTEQ2M parton 
         distribution function. The shaded bands indicate the uncertainty 
         on the predictions. }
\label{fig:xsec}
\end{figure}

The vector boson inclusive cross section times decay branching ratio 
follows from the number of background subtracted observed candidate 
events, corrected for efficiency, acceptance and luminosity: 
$    \sigma \cdot B \,=\,  { N_{obs} - N_{bkg}   \over 
                           {\rm A} \, \epsilon \, {\cal L} } 
$, 
where $N_{obs}$ is the observed number of events and $N_{bkg}$ the 
number of expected background events. $B$ indicates the 
branching ratio of the vector boson for the decay channel under study. 
The measured cross sections times branching ratio are listed in 
Table~\ref{table:xsec-result} and are compared with the theoretical 
prediction in Fig.~\ref{fig:xsec}. 
The theoretical predictions
for the total production cross section, 
calculated to ${\cal O}(\alpha_s^2)$ \cite{Neerven}, 
depend on three input parameters: 
the mass of the $W$-boson, $M_W$, 
the mass of the $Z$-boson, $M_Z$, 
and the structure of the proton. Using the \mbox{CTEQ2M} parton distribution 
functions~\cite{cteq}, 
the predictions for the total cross sections are 
$\sigma_W$ = 22.35~nb and $\sigma_Z$ = 6.708~nb. 
Using the leptonic branching ratio 
$B(W \rightarrow \ell\nu) = (10.84 \pm 0.02)\% $, 
as calculated following reference \cite{Rosner_w_width} using
$B(Z \rightarrow \ell\ell) = (3.366 \pm 0.006)\%$ as measured by the LEP 
experiments \cite{pdg}, 
the theoretical predictions for the total inclusive production 
cross section times branching ratio are 
$\sigma_W \cdot B(W \rightarrow \ell\nu)  = 2.42^{+0.13}_{-0.11}$~nb and 
$\sigma_W \cdot B(Z \rightarrow \ell\ell) = 0.226^{+0.011}_{-0.009}$~nb. 
The two largest uncertainties on the theoretical prediction are the 
choice of \pdf (pdf) (4.5\%) 
and the uncertainty due to using a NLO \pdf with a 
full ${\cal O}(\alpha_s^2)$ theoretical calculation (3\%).
The pdf uncertainty is not expected to decrease as rapidly as the statistical 
error on the measurement. 
The experimental error is dominated by the uncertainty on the luminosity. 
At the moment this uncertainty seems irreducible. Since there is good 
agreement between the theoretical prediction and the observed cross 
section it may therefore become advantageous to measure the luminosity 
using the observed $W$ event rate.

\begin{figure}[h]
    \epsfxsize = 7.0cm
    \centerline{\epsffile{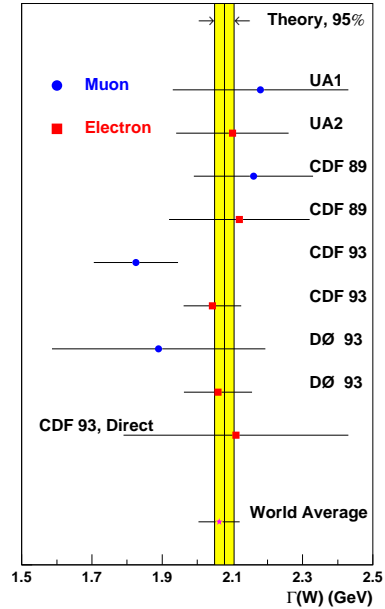}}
\caption{Measurements of $\Gamma_W$ compared with the $\SM$ 
         expectation. }
\label{fig:gamma_w}
\end{figure}

The ratio of the cross section measurements in which the error on the 
luminosity, common to both the $W$ and $Z$ events, completely cancels 
measures the leptonic branching ratio of the $W$-boson. 
It can be used, within the framework of the $\SM$, to extract
the total width of the $W$-boson: 

\begin{displaymath} 
    R \,=\,  { \sigma_W \cdot B(W\rightarrow\ell\nu)  \over
               \sigma_Z \cdot B(Z\rightarrow\ell\ell) } 
      \,=\,  { \sigma_W \over \sigma_Z } \cdot 
             {  \Gamma(W\rightarrow\ell\nu )  \over 
                \Gamma(Z\rightarrow\ell\ell ) } 
             {  \Gamma(Z) \over \Gamma(W) } 
\end{displaymath} 
which gives 
\begin{displaymath} 
    B^{-1}(W\rightarrow\ell\nu)  \,=\, 
            { \sigma_W \over \sigma_Z }                 \cdot 
            { 1        \over B(Z\rightarrow\ell\ell) }  \cdot 
            { 1        \over R }  \ .
\end{displaymath} 
Using the $\SM$ prediction for the partial 
decay width $\Gamma(W\rightarrow\ell\nu)$~\cite{Rosner_w_width}, 
the total decay width of the $W$, $\Gamma_W$, is given by  
\begin{displaymath} 
    \Gamma_W \,=\, 
                { \sigma_W \over \sigma_Z }     \cdot 
                { \Gamma(W\rightarrow\ell\nu)   \over 
                  B(Z\rightarrow\ell\ell) }     \cdot 
                { 1 \over R } 	\ .
\end{displaymath} 
The ratio of the cross sections, using the calculation of 
\cite{Neerven}, is determined to be 3.33 $\pm$ 0.03. 
Even though in the ratio the 
theoretical uncertainties also largely cancel, 
the error is still dominated by the choice of pdf's. 
Using, as before, the measured branching ratio 
$B(Z \rightarrow \ell\ell) = (3.367 \pm 0.006)\%$ and the 
theoretical prediction for the partial decay width 
$\Gamma(W\rightarrow\ell\nu)$ = 225.2 $\pm$ 1.5 MeV 
\cite{Rosner_w_width}  
the $W$ leptonic branching ratio, as determined from the 
combined \D0 electron and muon 
1992-1993 data, is (10.43 $\pm$ 0.44)\%; the CDF measured branching
ratio, based on the 1992-1993 electron data is 
(10.94 $\pm$ 0.33 $\pm$ 0.31)\%. Using the calculated 
partial leptonic branching ratio of the $W$, 
these measurements yield for the width 
$\Gamma_W~=~2.159~\pm~0.092$ GeV and 
$\Gamma_W~=~2.043~\pm~0.082$ GeV \cite{d0_cdf_xsec_1a}, respectively. 
The CDF value differs from their published value due to the use of more 
recent experimental measurements in evaluating the input parameters. 
% has been re-evaluated using the more recent input parameters quoted above. 
Figure~\ref{fig:gamma_w} shows the world $W$-width measurements together 
with the theoretical prediction 
\cite{d0_cdf_xsec_1a,ua1_gamma_w,ua2_gamma_w,cdf_xsec_89,cdf_gamma_w_direct}.

Taking into account that the ratio of the total cross sections
$\sigma_W /\sigma_Z$ is slightly different at a center of mass energy of 
630~GeV 
($\sigma_W /\sigma_Z (\sqrt{s}=630$ GeV) = 3.26 $\pm$ 0.09), 
and accounting for the 
correlation between the measurements at different center of mass energies 
through the choice of pdf's, the different values 
of $\Gamma_W$ can be combined to give a world average of 
$\Gamma_W = 2.062 \pm 0.059$ GeV, a measurement at the 3\% level. 
This is in good agreement with the $\SM$ prediction of 
$\Gamma(W) = 2.077 \pm 0.014$ GeV. The comparison of the measurement with 
the theoretical prediction can be used to set an upper limit 
on an \lq\lq excess width\rq\rq\ 
$\Delta \Gamma_W \equiv \Gamma_W{\rm (meas)} - \Gamma_W {\rm  (SM)}$,
allowed by experiment for  non--$\SM$ decay processes, such 
as decays into supersymmetric particles or into heavy quarks. 
Comparing the above world average value of $\Gamma_W$ with the 
$\SM$ prediction a 95\% C.L. upper limit of 
$\Delta\Gamma_W < 109$~MeV on unexpected decays can be set.

\section{Drell-Yan Production }

One of the unique features of \pbarp collisions is the intrinsic 
large range of 
available partonic center of mass energies. This allows for a study 
of the $Z$ line shape through the Drell-Yan process
($\qq \rightarrow (\gamma, Z \rightarrow ) \ \ell^+\ell^- )$ over 
a large di-lepton invariant mass region.
The low invariant mass region 
allows access to the small $x$ region of the parton distribution 
functions down to $x=0.006$, where $x$ is the fraction of the proton 
momentum carried by the parton. 
The high invariant mass region is populated by high $x$ partons and 
thus allows for a study of a possible substructure of the interacting 
partons. A possible substructure 
would manifest itself most prominently in a modification of the
$\gamma Z$ interference pattern whose effects are strongest in 
the region well above the $Z$ pole. 
Substructure of partons is most commonly
parametrized in terms of a contact interaction~\cite{contact},
\begin{equation}
    {\cal L} \,=\, {\cal L}_{\rm SM} \,+\,
                    \eta \, 
                   {g^2_0 \over {\Lambda^\eta_{ij}}^2 } \
                  (\overline{\psi}_i \gamma^\mu \psi_i ) \ 
                  (\overline{\psi}_j \gamma^\mu \psi_j ) 
\end{equation}
characterized by a phase, $\eta$, leading to constructive ($\eta=-1$) 
or destructive interference ($\eta=+1$) with the $\SM$ Lagrangian, and a 
compositeness scale, $\Lambda_\eta$, indicative of the 
energy scale at which substructure would be revealed. The indices 
$i,j$ refer to the chirality of the interacting fermions. 
The coupling constant $g^2_0$ is taken to be $4\pi$. 
By fitting the di-lepton invariant mass spectrum 
to various assumptions for the compositeness 
scale, phase of the interference and chirality of the interaction, 
lower limits on the compositeness scale can be set.

\begin{figure}[t]
    \epsfxsize = 8.0cm
    \centerline{\epsffile{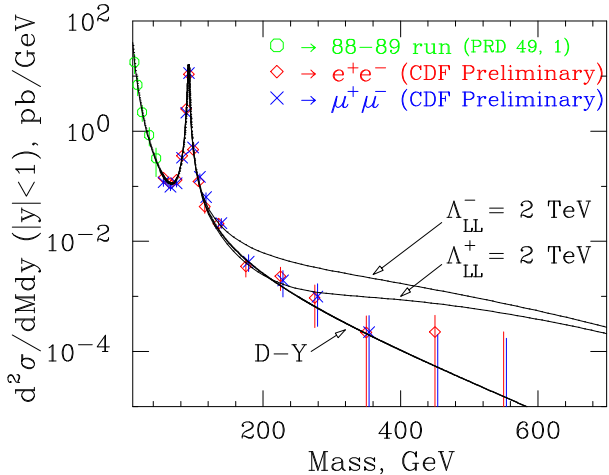}}
\caption{Double differential cross section $d^2\sigma / dM\,dy$ 
         for CDF electron and muon data combined. The open symbols 
         are from the 88--89 data. The solid symbols correspond to the 
         full Run I data. The curves are the theoretical predictions 
         for a lefthanded-lefthanded contact interaction with a scale 
	 of 2 TeV for constructive and destructive interference.  }
\label{fig:cdf_dy}
\end{figure}

The CDF experiment has measured the double differential Drell-Yan
cross section $d^2\sigma / dM\,dy$ for electron and muon pairs
in the mass range $11 < M_{\ell\ell} < 150$ GeV/c$^2$ for the
Run~Ia data~\cite{cdf_dy1a}, and
$40 < M_{\ell\ell} < 550$ GeV/c$^2$ for the Run~Ib data~\cite{cdf_dy1b}.
The di-electron invariant mass spectrum is measured over the rapidity 
interval $|\eta|<1$. Due to a more restricted coverage, the muon 
cross section has been determined only over the range $|\eta|<0.6\,$.
Figure~\ref{fig:cdf_dy} shows the measured cross section for electrons
and muons combined. 
The curves correspond to a leading-order calculation of the
Drell-Yan cross section with in addition a contact interaction between 
the quarks and leptons. The theoretical predictions have been normalized 
to the data over the $Z$ resonance region of $50 < M < 150$~GeV/c$^2$, 
which removes the uncertainty due to the luminosity and reduces the 
effect of the systematic uncertainty on the acceptance. 
Performing a maximum likelihood fit to the electron and muon data combined 
of Monte Carlo generated spectra for different assumptions for the contact 
interaction, scale factors are obtained as listed in 
Table~\ref{table:comp_limits}.
The limits imply that up to a distance of $10^{-17}$ cm the 
interacting particles reveal no substructure. 
The limits from other experiments on quark-lepton contact interactions 
are slightly less stringent. The lower limits range from 1.6 - 2.5 TeV 
from LEP~\cite{opal_contact}, and 1.0 - 2.5 TeV from HERA~\cite{h1_contact}.

\begin{table}[h]
\begin{center}
\begin{tabular}{l|ccccccc}
Model        & LL    & LR   & RL   & RR   & VV   &  AA   &  SC  \\ \hline
$\Lambda^+$  & 3.1   & 3.3  & 3.3  & 3.0  & 5.0  &  4.5  &  3.3	\\
$\Lambda^-$  & 4.3   & 3.9  & 3.7  & 4.2  & 6.3  &  5.6  &  3.3	\\
\end{tabular}
\end{center}
\caption[]{One-sided 95\% confidence level lower limits on the compositeness
	   scale (in TeV) for different chiralities of the contact 
	   interaction 
	   and different phases of interference with the SM Lagrangian. }
\label{table:comp_limits}
\end{table}

The distribution shown in Fig.~\ref{fig:cdf_dy} can be divided into two 
invariant mass regions: a pole region, $75 < M_{\ell\ell} < 105$~GeV/c$^2$ 
and a high mass region with $M_{\ell\ell} > $ 105 GeV/c$^2$ and the 
forward backward asymmetry, $\AFB$, can be measured for those two regions.   
$\AFB$ is defined as 
$\AFB = {\sigma_F \,-\, \sigma_B \over \sigma_F \,+\, \sigma_B } $ 
where $\sigma_{F(B)}$ is the cross section for fermion production in the
forward (backward) hemisphere. 
Because the left-handed and right-handed coupling of fermions to the 
$Z$ boson are not the same, the angular distribution of the 
outgoing fermion with respect to the incoming fermion in the 
parton center of mass frame exhibits a forward-backward asymmetry. 
Due to the changing polarization of the $Z$ boson as function of center 
of mass energy $\AFB$ has a strong energy dependence, which can be 
measured by studying $\AFB$ in different di-lepton invariant mass 
regions. 

Since 
the couplings of the fermions to the $Z$ boson depend on the fermion 
weak isospin and charge, $\AFB$ is different for different 
initial and final states. For the Drell-Yan process 
$\ppbar \rightarrow \ell^+\ell^- $ no distinction can be made between 
$\uu$ and $\dd$ initial states and therefore the asymmetry 
measured will be a convolution of both. It is interesting to note 
that this process is the time-reversal of the corresponding process 
at $e^+e^-$-machines and the measurements are complementary. 
At LEP and SLC the measurements are free from pdf uncertainties, whereas 
at the Tevatron, the light quark asymmetries are free from fragmentation 
uncertainties. 

The CDF experiment has measured $\AFB$ using the full Run~I data set
for di-electron final states with 
$|\eta_{\ell_1}| < 1.1 $ and $|\eta_{\ell_2}| < 2.4$~\cite{cdf_afb}.
The analysis yields 
$A_{FB} = 0.07 \pm 0.016 $ for 
75 $< M_{ee} < $ 105 GeV/c$^2$, and 
$A_{FB} = 0.43 \pm 0.10 $ 
for $M_{ee} > $ 105 GeV/c$^2$, compared to the $\SM$ predictions
of $A_{FB} = 0.054 \pm 0.001 $ and 
$A_{FB} = 0.528 \pm 0.006 $, respectively. 
Even though in the high mass region the asymmetry is measured with a
rather large error, these measurements still serve as a probe of 
extensions of the $\SM$ because models with additional heavy
neutral gauge bosons can substantially alter $\AFB$~\cite{rosner_zprime}

\section{$W$-mass }
\label{sec_wmass}

A possible choice of the fundamental parameters of the gauge sector 
of the standard model of electroweak interactions is the fine structure 
constant, $\alpha$, the Fermi constant, $G_F$ and the mass of the 
$Z$ boson, $M_Z$, all measured to very high precision. Within the $\SM$ 
the mass of the $W$ boson is then predicted and can be expressed in terms 
of these parameters. 
In the on-shell scheme the mass of the $W$ boson is given by 
\begin{equation} 
    M_W^2 \,=\, { M_Z^2 \over 2 }\, 
                \left( 1 \,+\, 
                       \sqrt{ 1 \,-\, {4\pi\alpha \over \sqrt{2} G_\mu M_Z^2 } 
                                \     {1 \over 1 - \Delta r } }
                \right) 
\label{eq:deltar}
\end{equation} 
where $\Delta r$ measures the higher order corrections. That is, 
at tree level $\Delta r$ vanishes. The dominant contribution to 
$\Delta r$ comes from the photon vacuum polarization which contributes 
about 0.06 to $\Delta r$. The other contributions come from the vector 
boson self-energies, the top quark which introduces a dependence quadratic
in $M_t$ and the Higgs boson, which adds a dependence logarithmic in $M_H$. 
Of course, also new physics would contribute 
to $\Delta r$. A precise measurement 
of the $W$ boson mass is thus a direct measure of the radiative corrections 
in the $\SM$ and combined with the measurement of $M_t$ it forms a constraint 
on the Higgs mass if the measurements are precise enough. In addition
it is sensitive to physics not included 
in the minimal $\SM$. 

In $W$ events produced in hadronic collisions in essence only two 
quantities are measured: the
lepton momentum and the transverse momentum of the recoil system. 
The latter consists of the ``hard'' $W$-recoil and the azimuthally 
symmetric underlying event contribution. It should be noted that the 
underlying event contribution is luminosity dependent.
The neutrino transverse momentum is equated 
to the total missing transverse energy in the event, $\etmisv$.
Since the
longitudinal momentum of the neutrino cannot be determined unambiguously, 
the $W$-boson mass is determined from the line shape in 
transverse mass, 
defined as 
\begin{equation} 
    m_T \,=\,  \sqrt{ 2\, p_T^\ell \, p_T^\nu \, 
                     (1 - \cos\varphi^{\ell\nu}) } \ .
\end{equation} 
Here $\varphi^{\ell\nu}$ is the angle between the lepton and neutrino in 
the transverse plane. 
Since there is no analytic description of the transverse mass distribution, 
the $W$-mass is determined by fitting Monte Carlo generated templates 
in transverse mass for different masses of the $W$-boson to the data. 
The distribution in $m_T$ exhibits a Jacobian edge characteristic
of two-body decays which contains most of the mass information. 
For the $W$-mass determination both the 
energy scale for the lepton and recoil system, which determine the 
peak position of the transverse mass distribution, as well as 
the resolution on the measured variables, which controls the
steepness of the Jacobian edge, are crucial. 

\begin{figure}[t]
    \epsfxsize = 7.0cm
    \centerline{\epsffile{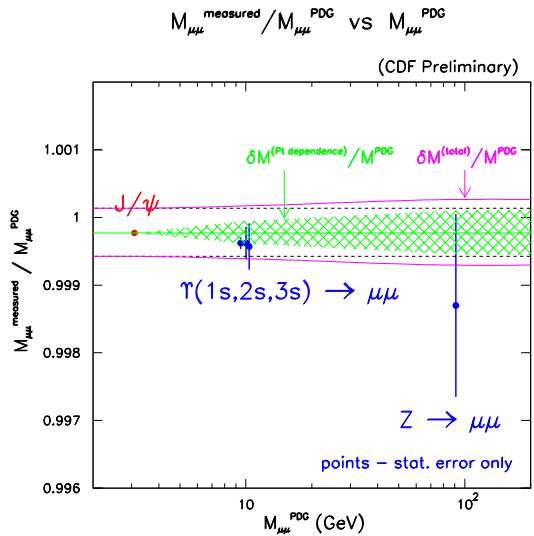}} 
    \vspace*{-2.0cm}
\caption{Ratio of measured and world average mass for various resonances. 
	 The dotted line indicates the total uncertainty on the momentum 
	 scale. }
\label{fig:p-scale} 
\end{figure}

Both the CDF and \D0 mass analyses discussed here are based on the Run~Ib
data, with the CDF analysis based on $W$ decays into muons and the 
\D0 analysis based on electrons. 
In the CDF $W$-mass analysis 
the momentum scale of the central
magnetic tracker is set by scaling the measured $J/\psi$-mass to 
the world average value using $J/\psi \rightarrow \mu^+\mu^-$ decays. 
Based on a sample of approximately 250,000 events the ratio of the measured 
and true $J/\psi$ mass has been determined to be 
0.99977~$\pm$~0.00048. 
The dominant contribution to the 
uncertainty on the momentum scale at the $J/\psi$ mass comes 
from the uncertainty in the amount of material the muons
traverse. To establish the momentum scale at the
$W$-mass the measured $J/\psi$-mass is studied as function of 
$\langle 1/p_T^2 \rangle$, extrapolated to zero curvature and verified 
with measurements of the $Z$ and $\Upsilon$ resonances. 
Figure~\ref{fig:p-scale} shows the measured momentum scale factor 
with its uncertainty as function of mass.  
The hatched region indicates the error incurred by extrapolating the 
momentum to the momentum scale appropriate for muons from $W$ decays. 
Also shown are ratios of the measured mass and world average mass 
for various other resonances. The measurements agree well, within the 
statistical uncertainty, with the scale determined from the $J/\psi$ mass.   
An overall uncertainty of 0.00048 on the momentum scale has been determined 
resulting in a 40~MeV/c$^2$ uncertainty on the 
$W$ mass.

At \D0 the $W$-mass is measured from $W\rightarrow e\nu$ decays. 
The electromagnetic (EM) energy scale is determined by calibrating to 
the $Z\rightarrow ee$ resonance in conjunction with the reconstruction 
of $\pi^0$ and $J/\psi$ decays.  Since the absolute energy scale 
of the EM calorimeter is not known with the precision required for this 
measurement, in essence 
the ratio of the $W$ and $Z$ masses is measured, anchored to the 
LEP $Z$ mass. 

To establish the energy scale 
it is necessary to determine to which extent a potential 
offset in the energy response, as opposed to a scale factor, is
responsible for the deviation of the ratio 
$M_Z^{\D0} \over M_Z^{\rm LEP}$ from unity. 
This was achieved by combining the measured $Z$ mass with the
measurements of 
$\pi^0 \rightarrow \gamma\gamma$ and $J/\psi \rightarrow e^+e^-$
decays and comparing them to their known
values.
If the electron energy measured in the calorimeter and its true energy
are related by
$E_{\rm meas} = \alpha \, E_{\rm true} + \delta$,
the measured and true mass values are, to first order, related
by $m_{\rm meas} = \alpha \, m_{\rm true} \,+\, \delta \, f $.
The variable $f$ depends on the decay topology and is given by
$f = {2(E_1 + E_2) \over m_{\rm meas} } \sin^2\gamma/2$,
where $\gamma$ is the opening angle between the two decay products
and $E_1$ and $E_2$ are their measured energies.
The ratio of the measured $W$ and $Z$ mass in this approximation can 
then be written as
\begin{eqnarray*}
    \lefteqn{
    \left. \frac{M_W (\alpha,\delta)}{M_Z (\alpha,\delta)}\right|_{\rm meas}
           =  }  \\
    & & 
    \left. \frac{M_W}{M_Z}\right|_{\rm true}
           \left[ 1 + \frac{\delta}{\alpha} \cdot
                      \frac{f_W \, M_Z - f_Z \, M_W}{M_Z \cdot M_W}
                      \right] \ . 
\end{eqnarray*}
It should be noted that the $W$ mass is insensitive to $\alpha$ if
$\delta=0$ and that the sensitivity to $\delta$ is proportional to $f$. 
Decays with different values for $f$ will thus have different 
sensitivity to the offset $\delta$. 
By combining the $Z\rightarrow ee$, 
$\pi^0 \rightarrow \gamma\gamma \rightarrow eeee$ and 
$J/\psi \rightarrow ee$ analysis, 
this in situ calibration of the EM calorimeter yields 
$\alpha = 0.95329 \pm 0.00077$ and 
$\delta =(-0.160  \pm 0.016 {}^{+0.060}_{-0.210})$~GeV. 
The asymmetric error on the offset is largely due to possible 
calorimeter nonlinearities, which is dominated by the 
uncertainty on the low energy response of the calorimeter. 
This uncertainty on the absolute energy scale
results, for the Run~Ib data sample, in an uncertainty on $M_W$ of 
70~MeV/c$^2$, of which 65~MeV/c$^2$ is due to the limited $Z$
statistics.

After the energy scale has been set, 
the $W$-mass is determined from a maximum likelihood fit 
of Monte Carlo generated templates in transverse mass to the 
data distribution. 
In the Monte Carlo model of $W$-production the triple differential 
production cross section is assumed to factorize into a term describing 
the mass dependence of the cross section and a term describing the 
longitudinal and transverse motion of the boson. 
The mass dependence is taken to be a 
relativistic Breit-Wigner resonance, adjusted for parton luminosity 
effects. 
The distribution in $p_T$ and rapidity of the $W$ boson is modeled 
according to the parametrization by 
Ladinsky and Yuan~\cite{ly} with a particular choice 
for pdf, thus including the 
correlation between the longitudinal and transverse momentum. 
The CDF choice for nominal pdf is the MRS\,R2- pdf~\cite{mrs_r2}, 
whereas \D0 uses the MRSA pdf~\cite{mrsa}.  
After the $W$ bosons are generated the decay is modeled, respecting the 
polarization of the boson~\cite{mirkes}. 
 
\begin{figure}[h]
    \epsfxsize = 5.0cm
    \centerline{\epsffile{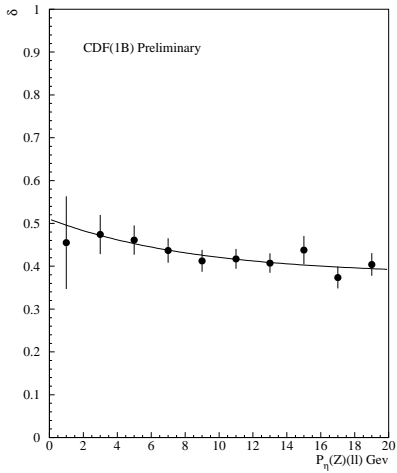}}
\caption[]{Deviation of the hadronic response relative to the  
	   electromagnetic response of the CDF calorimeter 
	   as function of  
	   $\vec{p}_T^{\,ee}$ projected onto the bi-sector of the two 
	   electron directions. }
\label{fig:cdf_recoil}
\end{figure}

The decay products are then traced through the detector and the detector 
response simulated. The parameters of the detector model 
are constrained by the data itself. 
The width of the $Z$-resonance, for example, provides a constraint on 
the momentum and energy resolution. The calorimeter response to the 
recoil of the $W$ boson is also determined using $Z$ events, 
by comparing the $p_T$ of the $Z$ obtained from
the two electrons, $\vec{p}_T^{\,ee}$, to that
obtained from the rest of the event,  $\vec{p}_T^{\,rec}$.
If the response of the hadronic calorimeter were equal to the response
of the EM calorimeter the vector sum of these two different measures of 
$p_T^Z$ would on average be zero. 
To minimize the contribution from the electron energy resolution,
the vector sum of these two quantities is projected
along the bisector of the two electron directions, called the 
$\eta$-axis. 
By studying 
$\vec{p}_T^{\,ee} \,+\, \vec{p}_T^{\,rec}$ 
as function of $\vec{p}_T^{\,ee}$ the relative response of the 
hadronic calorimeter with respect to the electromagnetic calorimeter 
is determined. Figure~\ref{fig:cdf_recoil} shows for the CDF 
experiment the quantity $\delta$ defined through the relation 
$|\vec{p}_T^{\,rec}| = (1-\delta) \, |\vec{p}_T^{\,ee}|$, 
as function of $\vec{p}_T^{\,ee} \cdot \hat{\eta} $. 
It can be seen that the CDF hadronic response is about 45\% lower than 
the electromagnetic response.

Both experiments model the underlying event using minimum bias data, 
mimicking the debris in the event due to spectator parton
interactions and the pile-up associated with multiple interactions,
and including the residual energy from previous beam crossings.

\begin{figure}[t]
\begin{center}
\begin{tabular}{c}
    \epsfxsize = 7.0cm
    \epsffile{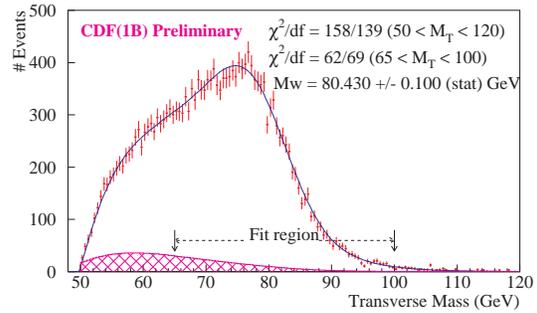}
\end{tabular}
\end{center}
\caption{CDF transverse mass distribution of $W\rightarrow \mu\nu$ decays 
         collected during the 1994-1995 run. The points are the data 
         and the line is the best fit. The arrows indicate the fit region. } 
\label{fig:mt_cdf} 
\end{figure}

The mass of the $W$ is obtained from a maximum likelihood fit in $m_T$ 
of events in the transverse mass range 
$ 65 < m_T < 100 $ GeV/c$^2$ ($ 60 < m_T < 90 $ GeV/c$^2$)
for CDF (\D0) to a data sample obtained by applying very stringent 
fiducial and kinematic cuts. Both experiments use only central leptons 
to determine the $W$ mass. 
Figure~\ref{fig:mt_cdf} 
shows the transverse mass distributions for the data
together with the best fit of the Monte Carlo 
for the Run Ib muon data for CDF. 
The $W$-mass is determined to be 
$M_W^e = 80.450 \pm 0.070 ({\rm stat.}) \pm 0.095 ({\rm syst.})$~GeV/c$^2$ 
by \D0 and 
$M_W^\mu = 80.430 \pm 0.100 ({\rm stat.}) \pm 0.120 ({\rm syst.})$~GeV/c$^2$ 
by CDF. 
Table~\ref{table:mw_sys} lists the systematic errors on the individual
measurements.

\begin{table}[h]
\begin{center}
\begin{tabular}{||l|r|r||} \hline\hline
                & \multicolumn{1}{c|}{ \ \ CDF \ }
                & \multicolumn{1}{|c||}{ \ \ \D0 \ }  \\
Source          & \multicolumn{1}{c|}{ $\mu$ }
                & \multicolumn{1}{|c||}{ $e$ }            
                \\ \hline
&&\\
Statistical                 & 100   &  70         \\
Energy/Momentum scale       &  40   &  70         \\
Other Systematics           & 115   &  70         \\ \hline 
\hspace*{0.4cm}
Lepton Angle                & ---   &  30         \\
\hspace*{0.4cm}
$e$ or $\mu$ resolution 
                            &  25   &  25         \\
\hspace*{0.4cm}
Recoil Model                &  90   &  40         \\
\hspace*{0.4cm}
$p_T^W$ Model, pdf's        &  50   &  25         \\
\hspace*{0.4cm}
QCD/QED corr's              &  30   &  20         \\
\hspace*{0.4cm}
$W$-width                   & ---   &  10         \\
\hspace*{0.4cm}
Backgrounds/bias            &  30   &  10         \\
\hspace*{0.4cm}
Fitting procedure           &  10   &   5         \\ \hline 
Total                       & 155   & 120         \\ \hline\hline
\end{tabular}
\end{center}
\caption[]{Errors on $M_W$ in MeV/c$^2$. }
\label{table:mw_sys}
\end{table}

\begin{figure}[h]
    \epsfxsize = 8.0cm
    \centerline{\epsffile{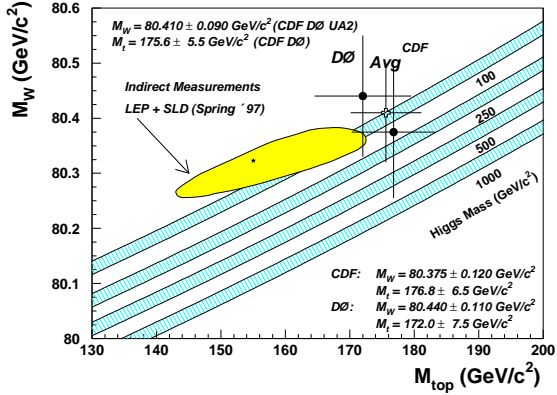}}
\caption{Relation between $M_W$ and $M_t$ in the $\SM$ for different values 
	 of $M_H$ together with the current direct and indirect 
	 measurements of $M_W$ and the measurements of $M_t$. }
\label{fig:mw_mt} 
\end{figure}

Combining
these measurements with previous $W$ mass measurements~\cite{mw_recent},
with a conservative assumption of a 50 MeV/c$^2$ correlated uncertainty 
due to the parton distribution functions and the input $p_T^W$ spectrum, 
gives a world average of $M_W = 80.410 \pm 0.090$~GeV/c$^2$ from the 
\pbarp collider experiments.  
Figure~\ref{fig:mw_mt} summarizes the current sta\-tus of the 
\pbarp measurements.

The expectation is that the theoretical uncertainties can be further 
constrained with a full analysis of the Run~I data. 
The uncertainty due to the pdf's can be 
constrained using the CDF measured $W$ charge asymmetry in conjunction 
with the world's data. 
The $W$ charge asymmetry is defined as 
\begin{displaymath} 
    A(y_\ell) \,=\,  
              { dN^+ (y_\ell) / dy_\ell \,-\, dN^- (y_\ell) / dy_\ell \over 
                dN^+ (y_\ell) / dy_\ell \,+\, dN^- (y_\ell) / dy_\ell } 
\end{displaymath} 
where $N^{+(-)}$ is the number of positively (negatively) charged leptons 
from the $W$ decay detected at pseudorapidity $y_\ell$. 
The rapidity distribution of the decay lepton is governed by both the 
$V-A$ structure of the $W$ decay and the longitudinal momentum 
distribution of the $W$ bosons. 
Since the $V-A$ structure of the $W$-decay is very well understood, 
the charge asymmetry measurement can then be used to probe the structure 
of the proton in the $x$ range 0.007 to 0.27$\,$. 

\begin{figure}[t]
    \epsfxsize = 6.0cm
    \centerline{\epsffile{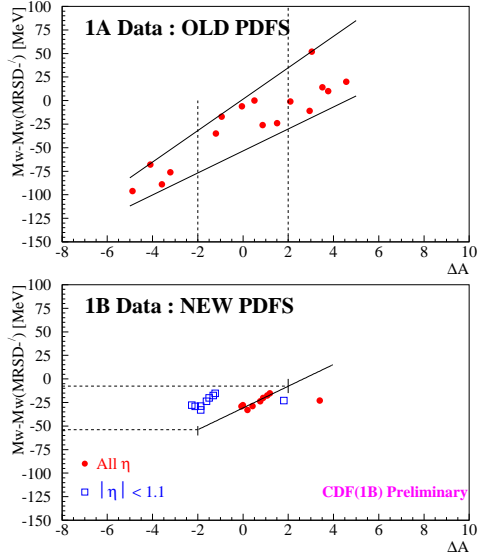}}
\caption{Change in $W$ mass versus significance of deviation between 
	 measured and predicted $W$ charge asymmetry for \lq\lq old\rq\rq\ 
	 (top) and \lq\lq new\rq\rq\ (bottom) pdf's. }
\label{fig:cdf_wasym} 
\end{figure}

CDF has updated the $W$ charge asymmetry 
measurement using the full Run~I data set with a
total integrated luminosity of 110~pb$^{-1}$. 
In addition, in the new analysis the rapidity coverage for muons 
has been extended by utilizing the 
forward muon toroids covering $1.95 < |\eta| < 3.6$, which collected
72~pb$^{-1}$ of data. The efficiency for electrons in the
plug calorimeter ($1.1 < |\eta| < 2.4$) was also substantially improved.
The agreement
between theory and experiment is quantified through the use of a 
significance parameter defined as 
\begin{equation} 
    \Delta A \,=\, { \overline{A}_{\,pdf} \,-\, \overline{A}_{\,data} 
		     \over 
                     \sigma(\overline{A}_{\,data})   }  \ \ .
\label{eq:sig_asym} 
\end{equation} 
where $\overline{A}$ is the weighted mean asymmetry 
as measured,  or as predicted by the DYRAD NLO Monte Carlo~\cite{dyrad}
using a particular set of pdf's.   
The top graph in Fig.~\ref{fig:cdf_wasym} shows the variation in $M_W$ 
versus $\Delta A$ for various \lq\lq old\rq\rq\ pdf's. Here 
\lq\lq old\rq\rq\ indicates pdf's determined without inclusion of the 
CDF Run~Ia $W$ charge asymmetry mea\-sure\-ment~\cite{cdf_wasym}. 
The label \lq\lq new\rq\rq\ refers to pdf's 
which do include that measurement in their determination 
of the structure functions. One anticipates a further reduction of 
the 25~MeV/c$^2$ uncertainty currently assigned due to the pdf  
uncertainty when the final $W$ charge asymmetry data is included in the 
global parton distribution fits. 

As a conservative estimate, 
currently also the uncertainty due to the input $p_T$ spectrum is taken 
as correlated between the two experiments.  The $W$ boson mass is extracted 
from events with low vector boson $p_T$, a region dominated by  
soft gluon emission. 
The soft gluon resummation formalism which describes this region contains 
a non-perturbative function parametrized in terms of phenomenological 
parameters, whose values have been derived from fits to Drell-Yan data. 
It is expected that the measured $p_T^Z$ spectrum will further  
constrain these parameters and thus the uncertainty due to the 
$W$ production model. 

The \pbarp average for $M_W$ can be combined with the LEP average of 
$M_W = 80.48 \pm 0.14$~GeV/c$^2$, the average of four measurements  
at both $\sqrt{s} = 162$ and 172~GeV~\cite{lep_mw}, and yields an average, 
assuming no correlation between the \pbarp and \ee measurements, of 
$M_W = 80.430 \pm 0.080$~GeV/c$^2$. 
The sensitivities of $M_W$ and $M_t$ to $\ln (M_H)$ are 
${\partial M_W \over \partial \ln (M_H) } \approx -75$~MeV and 
%per unit in $\ln(M_H)$  
${\partial M_t \over \partial \ln (M_H) } \approx  \ \ 12$~GeV 
per unit in $\ln(M_H)$. 
Given the current measurements with their uncertainties, 
reducing the uncertainty on $M_W$ by a factor of 2 
has about twice the power of an equivalent reduction of the top mass
uncertainty.

Looking at Fig.~\ref{fig:mw_mt} the standard observation is that the 
measurements agree with the $\SM$. Although such a statement is of course
correct, it does not really highlight the enormous achievement of these 
measurements of the $W$ mass. Recall that $\Delta r$ is a measurable 
quantity. 
The $\pbarp$ measurements alone of $M_W$ yield  
$    \Delta r  = 0.0335   \pm  0.0054 $. That is, $\Delta r$ is 
measured with a significance of $ 6.2\sigma $ from 
its tree level prediction. 
But, it is well known that $\Delta r$ is dominated by QED corrections. 
Following ref.~\cite{gambino_sirlin} 
the real electroweak bosonic corrections can be separated out by defining 
a $\Delta r_{res}$:  
\begin{equation} 
    {\alpha \over 1 \,-\, \Delta r  } 
    \,=\, 
    {\alpha(M_Z^2) \over 1 \,-\, \Delta r_{res}  } , 
\end{equation} 
where the pure QED corrections are absorbed in the running of $\alpha$. 
Using again the $\pbarp$ measurement of $M_W$ alone one finds 
$    \Delta r_{res}  \,=\, -0.0276  \pm  0.0059 $, 
a significance from the tree level prediction of $4.8\sigma $, 
a clear demonstration of weak bosonic 
corrections in the Standard Model. 
Including the LEP $M_W$ increases the 
significance to 5.5$\sigma$. 

An indirect measurement of the $W$-mass, through the measurement of the 
weak mixing angle $\sin^2\vartheta_W$, is described in detail in the 
contribution following this one~\cite{shaevitz}.

\section{Gauge Boson Pair Production }

The non-Abelian $SU(2)\times U(1)$ gauge symmetry of the $\SM$ implies 
that the gauge bosons self-interact. These self-interactions give rise 
to very subtle interference effects in the $\SM$. In fact, in the $\SM$  
the couplings are uniquely determined by the gauge symmetry in
order to preserve unitarity. 
An accurate measurement of the gauge boson self-interactions would 
constitute a stringent test of the gauge sector of the $\SM$ and any 
observed deviation of the couplings from their $\SM$ value would indicate
new physics. 

The formalism of effective Lagrangians is used to describe gauge boson 
interactions beyond the $\SM$. 
The most general effective electroweak Lagrangian 
contains $2\times7$ free parameters~\cite{Hagiwara}:
$	g_1^V, 	\kappa_V, 	\lambda_V,  
  	g_4^V, 	g_5^V,  
     	\widetilde{\kappa}_V, 	\widetilde{\lambda}_V , 
$
with $ V=\gamma,Z $. 
The parameter  
$       g_5^V, $ violates ${\cal C}$ and ${\cal P}$ but conserves ${\cal CP}$;
$ 	g_4^V, 	
        \widetilde{\kappa}_V$ and $ \widetilde{\lambda}_V$ 
violate ${\cal CP}$. 
In the $\SM$ 
$g_1^V=1, \kappa_V=1, $  and all other parameters vanish. 
For these two parameters one therefore introduces 
deviations from the $\SM$ values, 
$\Delta\kappa_V = \kappa_V - 1$ and 
$\Delta g_1^V   = g_1^V - 1 $.

Gauge boson self-interactions can be studied through di-boson production. 
The cross sections for di-boson production are generally rather small and 
a study of the full fourteen-dimensional parameter space is impossible. 
In general, two approaches are followed to reduce the parameter space. 
The \pbarp experiments generally set all parameters but two to their 
$\SM$ values and concentrate on 
$\Delta\kappa_V$, $\lambda_V$ because they have a direct physical 
connection through the magnetic dipole and electric quadrupole moment of 
the $W$ boson,
$    \mu_W = (e/2m_W)(1+\kappa_\gamma+\lambda_\gamma) $ and 
$    Q_W^e = (-e/m_W^2)(\kappa_\gamma-\lambda_\gamma) $~\cite{Kim}. 

The second approach, followed mainly by the LEP experiments, constructs
an effective Lagrangian with operators of higher dimension.
By imposing some restriction, like retaining only the lowest dimension 
operators, respecting 
${\cal C}, {\cal P}$ and ${\cal CP}$ invariance and requiring the 
Lagrangian to be invariant under 
$SU(2)\times U(1)$ and adding a Higgs doublet, the number of free 
parameters is reduced to just three~\cite{anomalous_lep}.
With further, rather ad hoc, requirements the parameter space can be 
reduced to just two free parameters~\cite{theorists}, with definite
relations between the different parameters~\cite{hisz}. 

If in the processes of di-boson production 
the couplings deviate even modestly from their 
$\SM$ values, the gauge cancellations are destroyed and a large increase 
of the cross section is observed. Moreover, the differential distributions 
will be modified.
%  giving rise to gauge bosons with a large transverse boost 
% since the largest gauge cancellations occur for highly boosted bosons. 
A $WWV$ interaction Lagrangian with constant anomalous couplings
would thus violate unitarity at high energies and therefore
the coupling parameters are modified to include
form factors~\cite{Baur}, that is, 
$\Delta\kappa (\hat{s}) = \Delta\kappa/(1+\hat{s}/\Lambda^2)^2 $ and 
$      \lambda(\hat{s}) =      \lambda/(1+\hat{s}/\Lambda^{2})^{2}$, 
where $\hat{s}$ is the square of the center of mass energy of the
subprocess. 
$\Lambda$ is a uni\-tarity preserving form factor scale 
and indicates the scale at which new physics would manifest itself. 
Limits on anomalous couplings are therefore always quoted for a particular 
value of $\Lambda$. 
In the next subsections some gauge boson pair production processes 
will be discussed.

\subsection{$WW \rightarrow \ell\ell^\prime\nu\nu^\prime$ Production }

Both the CDF and \D0 experiment have searched for $W$-boson pair production
$\pbarp \rightarrow WW + X \rightarrow \ell\ell^\prime\nu\nu^\prime$
$(\ell\ell^\prime = ee/e\mu/\mu\mu)$ based on 
data samples with an integrated luminosity of 108 and 97~pb$^{-1}$, 
respectively. The very few $\ttbar$ events recorded at the Tevatron are 
a background to this process, removed through cuts on the hadronic 
activity in the event. Both experiments observe 5 events over a background 
of $3.3 \pm 0.4$ and $1.2 \pm 0.3$ events for \D0 and CDF, respectively. 
This yields for CDF a mea\-su\-re\-ment of the cross section for 
$W$-pair production of 
$\sigma = 10.2^{+6.3}_{-5.1} \pm 1.6$~pb, to be compared to the 
$\SM$ prediction of 
$\sigma_{SM} = 9.5 \pm 2.9$~pb~\cite{cdf_ww_runI}.
It should be noted that the smallness of the cross section in 
itself is a beautiful demonstration of the gauge cancellations in 
the $\SM$. 

Since anomalous couplings not only result in an increase of the 
cross section but also significantly alter the differential distributions, 
limits on anomalous coupling parameters can be set by either using 
the event rate or by performing a fit to a differential distribution, 
generally taken to be the $p_T$ of one of the final state particles. 
Adopting the former approach, CDF has obtained the limits
$-1.1 < \Delta\kappa < 1.3 ~~(\lambda = 0)$ and 
$-0.8 < \lambda < 0.9 ~~(\Delta\kappa = 0)$ for 
$\Lambda = 1.0$~TeV. Performing a two-dimensional fit to the lepton 
$p_T$ spectra, \D0 obtained the limits 
$-0.62 < \Delta\kappa < 0.75 ~~(\lambda = 0)$ and
$-0.50 < \lambda < 0.56 ~~(\Delta\kappa = 0)$ for 
$\Lambda = 1.5$~TeV. Both sets of limits are obtained assuming 
$\Delta\kappa \equiv \Delta\kappa_{\gamma} = \Delta\kappa_Z$ and
$\lambda \equiv \lambda_{\gamma} = \lambda_Z$.

\subsection{$WW$ and $WZ$ Production }

Searches for particle production requiring two leptons in the 
final state always suffer in event rate due to the small leptonic 
branching ratios. 
When in the analysis described in the previous subsection only one
lepton is required, a substantial increase in event rate is obtained,
though at the cost of a much larger background. 
The background from $W/Z$+jet production to these processes 
is about 30 times higher than the signal production. 
Given the distinct characteristics of anomalous couplings
this background can be dealt with. 
Anomalous couplings modify the differential
distribu\-tions dramatically, especially the transverse momentum distribution 
of the $W$-boson. 
% The ratio 
% $\frac {\sigma_{WW}(p_T^W=200 \  {\rm GeV/c}) }
%        {\sigma_{WW}(p_T^W=20 \ {\rm GeV/c})}$ 
% is about $10^{-3}$, 
% whereas for only modest deviations from $\SM$ couplings
% ($\Delta\kappa = 0,\lambda=1.0$) 
% this ratio is about 0.5.  
By requiring the vector boson to have high transverse momentum 
the background is completely eliminated 
and a good sensitivity to anomalous couplings is retained. One 
completely loses sensitivity, however, to $\SM$ $W$-pair production. 

Both CDF and \D0 have looked for $W$-pair production using the leptonic 
decay of one of the $W$ bosons and the hadronic decay of the 
other~\cite{cdf_ww_wz,d0_ww_wz}.
Due to the limited jet energy resolution no distinction can be made 
between $WW$ and $WZ$ production and this analysis is thus sensitive to 
both processes. 
The jets from the hadronic decay of the $W$ or $Z$ boson are required 
to have an invariant mass consistent with the gauge boson mass, 
$60 < m_{jj} < 110$ GeV/c$^2$. 
Since no distinction can be made between $WW$ and $WZ$-production in this 
selection, CDF has increased the sensitivity of the study by including 
\pbarp$ \rightarrow WZ \rightarrow q{\overline q}^\prime\ell\ell$ events,
requiring the di-lepton invariant mass to reconstruct to the $Z$-boson mass.

\begin{figure}[h]
\begin{center}
\begin{tabular}{c}
    \epsfxsize = 6.0cm
    \epsffile{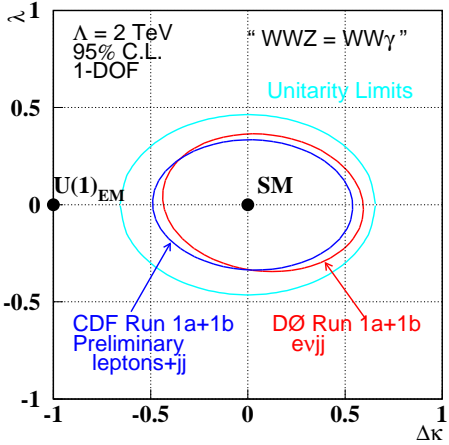}    
\end{tabular}
\end{center}
\caption{Contour limits on anomalous coupling parameters at the 95\% CL
         from CDF and \D0 from the analysis of $WW$ and $WZ$ production, 
	 using the hadronic decay of the $W$ or $Z$ boson, for a 
	 form factor scale of 2~TeV. 
	 The outer contour is the unitarity limit. }
\label{fig:wwljet} 
\end{figure}

Limits on anomalous couplings have been set by comparing the measured 
$p_T^W$ spectrum with the expectation using either the rate of events 
with $p_T^{W} > 200$~GeV/c, using both the electron and muon decays of 
$W$'s (CDF) or by performing a maximum likelihood fit to the full 
differential distribution in $p_T^W$ for $W$ decays into electrons
(\D0).
The 95\% CL contours in $\Delta \kappa$ and $\lambda$ obtained from this 
analysis for a form factor $\Lambda=2$~TeV are shown in 
Fig.~\ref{fig:wwljet}. 
The axis limits
% , the limits when only one coupling is 
% allowed to deviate from its $\SM$ value, 
are 
$-0.43 < \Delta\kappa < 0.59~~~(\lambda = 0)$ and 
$-0.33 < \lambda < 0.36~~~(\Delta\kappa = 0)$ for \D0, 
$-0.49 < \Delta\kappa < 0.54~~~(\lambda = 0)$ and 
$-0.35 < \lambda < 0.32~~~(\Delta\kappa = 0)$ for CDF, 
assuming again 
$\lambda_{\gamma} = \lambda_{Z}$ and 
$\Delta\kappa_\gamma = \Delta\kappa_{Z}$. 

Because $WW$ and $WZ$ production is sensitive to both the 
$WWZ$ and $WW\gamma$ coupling, 
information can be obtained on the $WWZ$ coupling alone by setting 
the $WW\gamma$ coupling to its $\SM$ value. The contour limits 
thus obtained show that a vanishing $WWZ$ coupling is excluded at 
a CL exceeding 99\%. 
As a matter of fact, this analysis is more sensitive to the $WWZ$
coupling than the $WW\gamma$ coupling due to the larger
coupling strength of the $WWZ$ vertex by a factor $\cot\vartheta_W$.

\subsection{$W\gamma$ Production }

In contrast to $\ee$ colliders where the $WW\gamma$ vertex 
cannot be separated from the $WWZ$ vertex, $\pbarp$ colliders allow for
the study of the $WW\gamma$-vertex only, without interference from the 
$WWZ$ coupling, through the study of $W\gamma$ production. 
In this analysis one searches for photons produced in association with 
a $W$ boson. 
% Most photons produced in association with a $W$, however, are radiated
% off the initial or final state fermion. The only channel that allows for
% a direct probe of the triple gauge boson vertex is the 
% $s$-channel contribution of a photon radiated from a $W$. 
In these analyses $W\gamma$ events are selected by requiring, 
in addition to the regular 
$W$ selection criteria, an isolated photon with high transverse energy 
% $E_T^\gamma > 10\ (7)$ GeV for \D0 (CDF) 
in the central pseudo-rapidity range 
$|\eta_\gamma | < 1.1$ for CDF and with 
$|\eta_\gamma | < 1.1$ or $1.5 < |\eta_\gamma | < 2.5$ for \D0. 
To reduce the contribution from radiative events the photon is 
required to be well separated from the lepton from the $W$-decay.

\begin{figure}[h]
    \epsfxsize = 8.0cm
    \centerline{\epsffile{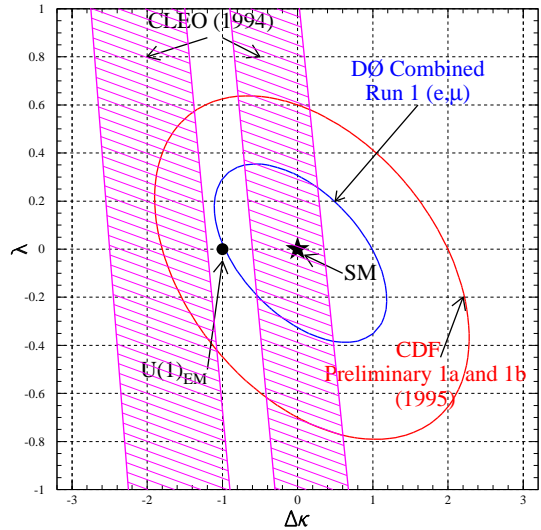}}
\caption{Limits on anomalous $WW\gamma$ couplings from $W\gamma$ analyses. 
	 The shaded bands are the constraints from CLEO. }
\label{fig:wgamma_lim}
\end{figure}

The dominant background to this process is
$W$+jet production with the jet being identified as a photon in the 
detector. For both experiments the signal to background ratio is 
about 3:1 with a probability of a jet \lq\lq faking\rq\rq\ a photon 
of about $10^{-3} - 10^{-4}$. 
The observed number of events are in good agreement with the number 
of events expected from $\SM$ processes and from the different 
background sources.  Limits on anomalous couplings are set by performing 
a maximum likelihood fit to the observed $p_T^\gamma$ spectrum. 
Based on a partial data set CDF obtained the limits: 
$-1.8 < \Delta\kappa < 2.0~~~(\lambda = 0)$ and 
$-0.7 < \lambda < 0.6~~~(\Delta\kappa = 0)$. The \D0 experiment has 
finalized the analysis using the full Run~I data sample and obtains 
$-0.93 < \Delta\kappa < 0.94~~~(\lambda = 0)$ and 
$-0.31 < \lambda < 0.29~~~(\Delta\kappa = 0)$~\cite{d0_wgamma_run1}. 
Both sets of limits are obtained for 
a form-factor scale of 
$\Lambda=1.5$~TeV (see Fig.~\ref{fig:wgamma_lim}).

The decay rate for $b\rightarrow s\gamma$ can also be used to set limits
on anomalous couplings since the process is sensitive to photon radiation off
the $W$-boson in the penguin diagram. The branching ratio has been measured
by CLEO to be 
$B(b\rightarrow s\gamma) = (2.32 \pm 0.57 \pm 0.35)\, 10^{-4}$ 
\cite{cleo_bsgamma}. The upper limit on this branching ratio excludes the 
outer regions in Fig.~\ref{fig:wgamma_lim}. 
The narrow region between the two
allowed CLEO bands is excluded by the lower limit.

\subsection{Combined Result on $WW\gamma$ Coupling }

The studies of $W\gamma$ and $WW/WZ$ production can be
combined to improve on the limits on anomalous couplings. 
When combining results, the correlation between the different 
analyses needs to be addressed. 
Some of the dominant common systematic uncertainties are 
due to the method of estimating the background and 
the uncertainty in structure functions and photon identification. 
The \D0 experiment has carried out a combined fit to the three
data sets corresponding to the 
$WW$, $WW/WZ$ and $W\gamma$ analyses based on the full  Run~I data.  
The significantly improved preliminary limits are: 
\begin{center}
\begin{tabular}{cc}
    $-0.33 < \Delta\kappa < 0.45$    & $(\lambda = 0)$         \\
    $-0.2 < \lambda < 0.2$           & $(\Delta\kappa = 0)$,
\end{tabular}
\end{center}
where it was assumed that the $WWZ$ couplings and the $WW\gamma$
couplings were equal. 
Strong contraints are anticipated when all results from both CDF 
and \D0 are combined.

As for the gauge boson self-interactions, again, agreement with the 
$\SM$ expectation is observed. 
But, deviations were not really expected, since most models predict 
anomalous couplings of order ${\cal O}(M_W^2/\Lambda^2)$. 
These measurement, however, do tell us that a $W$ boson is more 
than just an electrically charged boson. 
The point labeled $U(1)_{\rm EM}$ in 
Figs.~\ref{fig:wwljet} and~\ref{fig:wgamma_lim} 
corresponds to the values of the couplings if the $W$ boson would 
only couple electromagnetically. It is clear that this point is excluded 
by the data, showing that the $W$ boson is really a gauge boson.

\subsection{$Z\gamma$ Production }

The $ZZ\gamma$ and $Z\gamma\gamma$ trilinear gauge
boson couplings are described in a way analogous to the 
$WWV$ couplings. 
These couplings, absent in the $\SM$, are suggested by some theoretical
models which imply new  physics. 
The most general Lorentz and gauge invariant $ZV\gamma$ vertex is 
described by eight coupling  parameters,  $h^V_i,~(i=1...4)$, 
where $V = Z,\gamma$, 
which also are modulated by form factors to preserve 
unitarity,
$h^V_i =  h^V_{i0} / (1 + \hat{s}/\Lambda^2)^n$, 
where  $\hat{s}$ is the square of the invariant mass
of the $Z\gamma$ system and $\Lambda$ is the form-factor scale. 
The energy dependence of the form factor is assumed to be 
$n = 3$ for $h^V_{1,3}$  and $n = 4$ for $h^V_{2,4}$~\cite{Baur_Z}. 
Such a choice yields the same asymptotic energy behavior for 
all the couplings.

The study of anomalous couplings in the process 
$Z\gamma\rightarrow \ell\ell\gamma$ is analogous to the 
$W\gamma$ analysis, that is, events are 
selected with a photon produced in 
association with a $Z$ boson and the observed number of events compared 
with the number of expected radiative $Z$ and  background 
events~\cite{cdf_zgamma,d0_zgamma}. 
The $p_T^\gamma$ spectrum is again used to set limits on possible anomalous 
couplings.

The \D0 experiment has recently performed a new analysis based on the 
Run~Ia data, looking for the
decay $Z\gamma\rightarrow \nu\nu\gamma$~\cite{d0_zvv_1a}. 
This channel has previously been studied only in 
$e^+e^-$-collisions~\cite{l3_zgamma,delphi_zgamma}. 
Sensitivity to anomalous couplings in this channel is much  
higher than in the di-lepton decay modes due
to the higher decay rate into neutrinos and the absence  
of the radiative $Z$ decay background. 
The overall background, however, is still extremely high, leading 
to very stringent event selection criteria. 
To reduce the background from $W$+jet events 
with the electron or jet being misidentified as a photon 
the $E_T^\gamma$ and \etmis were required to exceed 40~GeV. 
In addition, events with at least one jet with 
$E_T^j > 15$~GeV were rejected. 
The  remaining  background  was dominated  by cosmic rays and muons from
beam halo which radiated in the calorimeter. 
This background  was suppressed  by rejecting events with a 
reconstructed muon or a minimum ionizing trace 
in the  calorimeter close to the photon cluster.  
The residual background, which had roughly equal contributions from 
$W \to e\nu$ decays and muon bremsstrahlung, was derived from data.

\begin{figure}[h]
    \epsfxsize = 8.cm
    \centerline{\epsffile{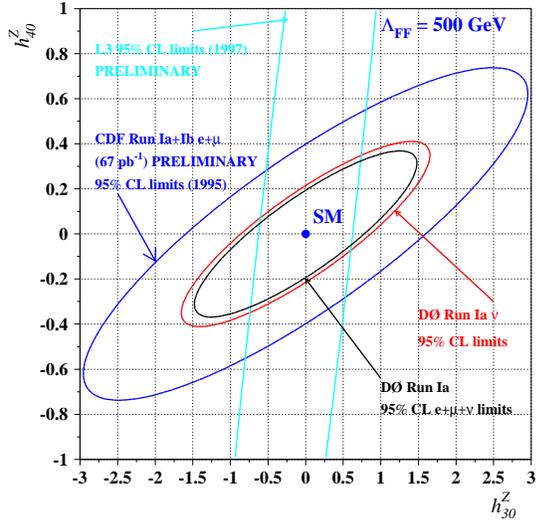}}
\caption{Limits on anomalous ${\cal CP}$-conserving $ZZ\gamma$ couplings from
         $Z(\ell\ell)\gamma$ and $Z(\nu\nu)\gamma$ production 
         for a form-factor scale $\Lambda = 500$~GeV.}
\label{fig:zg_limits}
\end{figure}

Four candidate events are observed on an expected background of  
$5.8 \pm 1.0$  events and a $\SM$ prediction of $1.8 \pm 0.2$ events. 
Although the signal-to-background ratio is less than one, 
the sensitivity to anomalous couplings is still high,
since the background is concentrated at low $E_T^\gamma$ while
the anomalous coupling contribution is almost flat in $E_T^\gamma$ up to
the kinematic threshold of the process.  
Limits on anomalous couplings were  set at 95\%  CL by a fit to 
the  $E_T^\gamma$ spectrum and gives 
$|h^Z_{30}| < 0.87$,  $|h^Z_{40}| < 0.21$ for $\Lambda = 500$~GeV. 
These limits, based on 14~pb$^{-1}$ of data, are more stringent than the
limits obtained from the analysis of the full Run~I data using the electron 
and muon decays of the $Z$ boson, indicating the strength of the neutrino 
channel. 
A summary of all the limits is shown in Fig.~\ref{fig:zg_limits}.

\section{Conclusions}
A wide variety of properties of the $W$ and $Z$-bosons are now being 
studied at hadron colliders with ever increasing precision, at the 
highest energy scales achievable. 
% The mass of the $W$-boson is now measured to about 0.2\%. 
All results, including the results from $\ee$ 
colliders~\cite{stickland}, are in good agree\-ment with the 
$\SM$. 
It is widely anticipated, though, that the $\SM$ is just an 
approximate theory and should eventually be replaced by a more 
complete and fundamental description of the underlying forces in nature. 
With the new data from LEP~2, SLD and the Tevatron, 
and with the planned upgrades of the accelerators as well as the
experiments, the projected uncertainties
on some fundamental parameters, especially the $W$ mass,  
should provide the tools to take another ever more critical 
look at the $\SM$, without any theoretical prejudice.

\section{Acknowledgements}
I would like to thank Debbie Errede, Bob Wagner and Darien Wood who have been 
very cooperative and the organizers for a very stimulating conference 
in a splendid setting.

\end{document}